\documentclass[aps, pra, preprint, superscriptaddress]{revtex4-2}
\usepackage{float}
\usepackage[utf8]{inputenc}
\usepackage[]{graphicx}
\usepackage[english]{babel}
\usepackage{amsmath,amsfonts,amssymb}
\usepackage{xcolor}
\usepackage{booktabs}
\usepackage{lineno}
\usepackage{textcomp, gensymb}
\usepackage[euler]{textgreek}
\usepackage[version=4]{mhchem}
\usepackage{geometry}
\usepackage{array}
\usepackage{boldline} 
\usepackage{float} 
\usepackage{tcolorbox} 

\makeatletter

\newenvironment{Figure}{%
\par\addvspace{12pt plus2pt}%
\def\@captype{figure}%
}{%
\par\addvspace{12pt plus2pt}%
}%

\long\def\@makecaption#1#2{%
\vskip\abovecaptionskip
\sbox\@tempboxa{#1: #2}%
\ifdim \wd\@tempboxa >\hsize
#1: #2\par
\else
\global \@minipagefalse
\hb@xt@\hsize{\hfil\box\@tempboxa\hfil}%
\fi
\vskip\belowcaptionskip}

\makeatother

\DeclareUnicodeCharacter{202F}{+}
\DeclareUnicodeCharacter{2009}{+}
 
\begin{document}

\title{Synthesis of Y$_3$Fe$_4$H$_{20}$ as a new prototype structure for ternary superhydrides recoverable at ambient pressure}

\author{Maélie Caussé}
\email{Present address: Department of Materials Science and Metallurgy, University of Cambridge, 27 Charles Babbage Road, Cambridge, CB30FS, United Kingdom}
\affiliation{CEA DAM, DIF, F-91297 Arpajon, France}
\affiliation{Université Paris-Saclay, CEA, Laboratoire Matière en Conditions Extrêmes, 91680 Bruyères-le-Châtel, France}

\author{Loïc Toraille}
 \email{author to whom correspondence should be addressed: loic.toraille@cea.fr}
\author{Grégory Geneste}
 \email{author to whom correspondence should be addressed: gregory.geneste@cea.fr}
\author{Paul Loubeyre}
\affiliation{CEA DAM, DIF, F-91297 Arpajon, France}
\affiliation{Université Paris-Saclay, CEA, Laboratoire Matière en Conditions Extrêmes, 91680 Bruyères-le-Châtel, France}


\begin{abstract}

Reaching pressures in the 100~GPa range enables the synthesis of hydrogen-rich compounds, with nontraditional H stoichiometries and H sublattices, called superhydrides. Record-breaking superconductivity temperature in some superhydrides have attracted great interest. A crucial next step is to stabilize superhydrides outside of high-pressure environments, leading to a search beyond binary hydrides to ternary hydrides. Here, we report the synthesis of Y\(_3\)Fe\(_4\)H\(_{20}\) at pressures starting at 60~GPa by compressing an hydrogenated Y-Fe compound, embedded in hydrogen in a laser-heated diamond anvil cell. Single-crystal X-ray diffraction allowed us to resolve the Y\(_3\)Fe\(_4\) lattice skeleton, and a constrained \textit{ab initio} structural search was used to position the hydrogen atoms. FeH\(_8\) cubic molecular units form building blocks which are connected edge-to-edge by sharing two hydrogen atoms, creating a framework that hosts Y cations. Remarkably, Y\(_3\)Fe\(_4\)H\(_{20}\) maintains its structure through decompression, making it the first superhydride recovered metastable under ambient conditions. 

\end{abstract}

\maketitle

\section*{Introduction}

The H atom bonding flexibility in the solid state leads to various types of hydrides at ambient pressure, classified as interstitial, ionic, complex, and covalent, with textbook examples being PdH, LiH, Mg\(_2\)FeH\(_6\), and AlH\(_3\), respectively. Fifteen years ago, a new chemistry of hydrogen with metals was projected to take place in the 100~GPa range~\cite{Zurek2009} which has launched an extensive exploration for H-rich compounds formed by combining H with another element of the periodic table. Many compounds with high H concentration were calculated to be stable, and these have been called superhydrides (see reviews in ref.~\cite{FloresLivas2020,Du2022,Sun2023}). FeH\(_5\)~\cite{Pepin2017} and LaH\(_{10}\)~\cite{Geballe2018} were the first superhydrides to be synthesized. Superhydrides are characterized by a hydrogen concentration higher than that of the stable hydrides of the same metal at ambient pressure, and a diversity of dense hydrogen sublattices with nearest H-H distances smaller than in conventional hydrides ($<$ 1.5~Å around 100~GPa excluding the presence of H\(_2\) molecular units, see the Supplementary Material). In LaH\(_{10}\), the H atoms form an extended weakly bonded clathrate structure, with nearest H-H distances around 1.1~Å, whereas in FeH\(_5\) four planes slabs of atomic hydrogen are bonded to iron atoms. The various known functionalities of ambient pressure hydrides, such as superconductivity, hydrogen storage, and hydridic ionic conductivity, are magnified for superhydrides. This is exemplified in the case of LaH\(_{10}\) which holds a record superconductivity temperature of 262~K~\cite{Somayazulu2019, Drozdov2019} and exhibits a superionic hydridic state~\cite{Causse2023}. Unfortunately, none of these superhydrides have yet been stabilized under ambient conditions, preventing any potential societal applications.

The possibility of recovering these dense hydrogen arrangements, typical of superhydrides, in a compound under ambient conditions is an open question. Almost all possible binary superhydrides have been tried out, so the search is currently targetting ternary systems, which offer more structural prototypes and potential for stability at lower pressure. Computational \textit{ab initio} random searches to systematically screen for stable ternary superhydrides are a formidable challenge. Various routes have been followed recently to reduce the phase space to be explored. One such way is to test for chemical doping by an electron donor element, as in the case of the prediction of Li\(_2\)MgH\(_{16}\)~\cite{Sun2019}. Another method is to look for substitutions of two elements in a given superhydride structure, as observed in (La,Y)H\(_{10}\) and (La,Ce)H\(_9\)~\cite{Semenok2021, Bi2022}. Another approach is to assemble a combination of heavy X and light Y elements with disparate sizes, which makes a very effective chemical compression originating from the X-Y scaffolding. A ternary XYH\(_8\)-type structure was predicted~\cite{Zhang2022} and recently synthesized in LaBeH\(_8\), which shows superconductivity above 100~K at sub-megabar pressures~\cite{Song2023}. LaBeH\(_8\) can be viewed as a saltlike configuration of La\(^{2+}\) BeH\(_8\)\(^{2-}\)~\cite{Wang2024}, yet it is not stable at ambient pressure. Finally, two works have recently performed high-throughput computation for ternary hydrides, over many combinations in the periodic table, to explore the possibility of high-temperature ambient-pressure hydride superconductivity. Their most promising candidate was Mg\(_2\)IrH\(_6\), with a critical temperature $T_c$ of 160~K~\cite{Dolui2024, Sanna2024}. It is a complex hydride based on the [IrH\(_6\)]\(^{3-}\) hydrido cluster, not experimentally observed yet~\cite{Hansen2024}. To keep this computation feasible, calculations were limited to high-symmetry crystal structures with fewer than 20 atoms per unit cell \cite{Dolui2024}, which likely explains why no superhydride have been found under these constraints.

Here we investigate the synthesis of ternary superhydrides in the Y-Fe-H system, without pre-guidance from calculations. We used the recently developed DAFi tool~\cite{Aslandukov2022}, which is revolutionizing high-pressure chemistry, to perform single-crystal refinement of micron-sized crystals powder obtained after chemical reaction in a laser-heated diamond anvil cell. Some chemical ingredients have guided our selection of the Fe-Y-H system, in particular the already observed high uptake of hydrogen under pressure by both Fe and Y elements and by their alloy, the YFe\(_2\) Laves phase. Fe and Y form binary superhydrides in the 100~GPa range, FeH\(_5\)~\cite{Pepin2015} and YH\(_{10}\)~\cite{Kong2021}, with very different H sub-lattices: 2D atomic metal slabs and H clathrate respectively. When compressing the stable compound YFe\(_2\)H\(_{4.2}\) in hydrogen without heating, YFe\(_2\)H\(_7\) was synthesized above 20~GPa reaching then a limit of insterstitial hydrogen uptake~\cite{Causse2025}. Finally, [FeH\(_6\)]\(^{4-}\) is a well-known hydrido cluster which is a building block of ambient pressure complex hydrides. It has been recently shown that under application of hydrogen pressure in the 10~GPa range, a novel homoleptic pentagonal bipyramidal [FeH\(_7\)]\(^{3-}\) is formed which can be the building block of novel complex transition metal hydrides~\cite{Spektor2020}. Therefore, the question arises: why not stabilizing [FeH\(_8\)]\(^{2-}\) by going to higher pressure? Using a laser-heated diamond anvil cell, we synthesized a ternary superhydride, Y\(_3\)Fe\(_4\)H\(_{20}\), at pressures as low as 60~GPa. Its structure reveals a novel hydride structure type, with chains of [FeH\(_8\)] moieties encapsulating Y cations, which can be seen as the analogue of a silicate-like structure for oxides. Remarkably, this superhydride remains metastable at ambient conditions.

\section*{Results}
\subsection*{Y$_3$Fe$_4$ structure resolution by single-crystal diffraction}
The high pressure hydrogen environment on YFe$_2$ was created by loading micron-sized samples of YFe$_2$H$_{4.2}$ in excess hydrogen into the sample chamber of Diamond Anvil Cells (DAC). Three DACs were loaded (run1, run2 and run3), some with two samples, giving a total of five different syntheses which demonstrate excellent repeatability in the results. As shown in Fig.~\ref{fig:1}, the samples were laser-heated at different pressures which enabled to overcome kinetic barriers in the transformation towards the most stable compounds in the Y-Fe-H system at a given pressure. The samples were characterized by single-crystal X-Ray-Diffraction (XRD) and angular-dispersive powder XRD at the ID15b and ID27 beamlines of the ESRF synchrotron. YFe$_2$H$_{7}$ was observed as the inital state of our high-pressure samples before the laser heating was performed, as expected based on previous work~\cite{Causse2025}. Under laser heating at 60~GPa (run2), 80~GPa (run3) or 83~GPa (run1), a clear sample recrystallisation took place, as depicted in Fig.~\ref{fig:1}. The samples chemically reacted to form high-quality micron-sized crystals. This allowed to perform a full single-crystal structure solving and refinement by combining the softwares CrysAlisPro~\cite{Crysalis}, DAFi~\cite{Aslandukov2022} and Jana2006~\cite{Petricek2014}. The refinement of the dominant phase yielded an orthorhombic unit cell $Cmcm$ of 4 formula units, with the formula Y$_3$Fe$_4$ (a=3.8312~\AA, b=9.9016~\AA, c=13.2019~\AA~and V=500.8142~\AA$^3$ at 83~GPa) and the Wyckoff positions of yttrium and iron atoms are respectively 4c and 8f for Y, 4b, 8f and 4c for Fe (see the Supplementary Material). Due to the low scattering power of hydrogen atoms, it was not possible to experimentally determine the hydrogen positions in that structure and consequently the H content, so the complete formula of this new compound was an indeterminate Y$_3$Fe$_4$H$_x$ at this stage. The DAFi analysis enabled us to identify multiple domains of this new compound with different orientations present in the sample, and X-ray mapping allowed us to reconstruct its volume extent. Another minor phase was also present and its structure Y$_2$Fe$_3$H$_{x}$ is given in the Supplementary Material. Since Y$_3$Fe$_4$H$_x$ is the dominant phase and was subsequently recovered stable at ambient pressure, it is the focus for our more detailed characterization, as reported below.

\subsection*{Y$_3$Fe$_4$H$_{20}$ structure resolution by constrained AIRSS}

The number of H atoms and their positions in the Y$_3$Fe$_4$ lattice were obtained at 90~GPa using first-principles density-functional theory (DFT) calculations under a constrained random search based on the AIRSS algorithm~\cite{Pickard2011}, as described in the methods section. Essentially, this involved fixing the experimentally resolved $Cmcm$ space group, the atomic positions and the lattice parameters of the Y$_3$Fe$_4$ skeleton (in the generation of the random structures, these positions and lattice parameters being subsequently structurally optimized). Furthermore, the measured volume of the unit cell, 500~\AA$^3$, should not be too different from the volume obtained by assuming additivity of the volume of the stable hydrides of the two end-member elements, YH$_x$ and FeH$_x$, stable around the same pressure, hence a mixture of YH$_3$ and FeH$_2$, or YH$_3$ and FeH$_3$. Correspondingly, the random searches for hydrogen content were constrained to vary between 17 and 21 per formula unit. The formation enthalpies of the various stable compounds obtained for each stoichiometry were then plotted and compared to the convex-hull, which shows thermodynamically stable compounds regarding decomposition towards H$_2$ and Y$_3$Fe$_4$H$_{17}$, as shown in Fig.~\ref{fig:2}. Two stable compounds Y$_3$Fe$_4$H$_{18}$ and Y$_3$Fe$_4$H$_{20}$ were good candidates. The hydrogen-excess experimental conditions ensure that the most stable compound to be synthesized should fall on the convex-hull with the highest H concentration, hence the selection of Y$_3$Fe$_4$H$_{20}$. Additionally, we performed enthalpy calculations of all possible simple compounds which Y$_3$Fe$_4$H$_{20}$ could decompose into (Y, Fe, H$_2$, YH$_x$, FeH$_x$, YFe$_2$H$_x$), at pressures ranging from 90 to 0~GPa. The results, including a total of 48 possible reactions, are gathered in the Supplementary Material: for all pressures studied~$\geq$~15~GPa, the reactions of formation of Y$_3$Fe$_4$H$_{20}$ from the simpler compounds listed above, all have a negative enthalpy of reaction, 
indicating that Y$_3$Fe$_4$H$_{20}$ is energetically more favorable with respect to its decomposition into simpler compounds. At 0~GPa, however, the calculations indicate that Y$_3$Fe$_4$H$_{20}$ is unstable with respect to its decomposition for two cases: to form a mix of Fe, H$_2$ and YH$_3$, as well as a mix of FeH, H$_2$ and YH$_3$. This is expected since Fe, FeH, H$_2$ and YH$_3$ are all compounds observed stable at ambient pressure.
The dynamical stability of Y$_3$Fe$_4$H$_{20}$ was verified at 90~GPa by computing the phonon dispersion curves using the Density Functional Perturbation Theory (DFPT). Phonon calculations, including the phonon dispersion curves and corresponding vibrational densities of states, are shown in Fig.~\ref{fig:2} and indicate that the Y$_3$Fe$_4$H$_{20}$ structure is dynamically stable at 90~GPa, as evidenced by the absence of imaginary phonon modes. 

Our DFPT analysis reveals a clear separation between the dynamics of the heavy Y and Fe atoms and that of the lighter H atoms (Fig.~\ref{fig:2}). Overlapping peaks in the vibrational density of states for the metallic atoms Y and Fe are observed. Additionally, the H-related phonon dispersion curves show a weak dispersion of the phonon modes, which is unusual. 

The electronic densities of states and electronic band structure of Y$_3$Fe$_4$H$_{20}$ computed at 90~GPa are shown, in Fig.~\ref{fig:2} and in the Supplementary Material respectively. From these, we can conclude that Y$_3$Fe$_4$H$_{20}$ is metallic. However, this superhydride is unlikely to be a high-T$_c$ superconductor, as it exhibits a low H-projected electronic density of states at the Fermi level. 

The crystal structure of Y$_3$Fe$_4$H$_{20}$ is depicted in Fig.~\ref{fig:2}. It is remarkable to see that the structure is formed of molecular [FeH\(_8\)] entities, while not being a complex hydride. The [FeH\(_8\)] entities are connected to form chains encapsulating the Y cation, which is a previously unobserved structure for an H compound.

\subsection*{Equation of state of Y$_3$Fe$_4$H$_{20}$}
Single-crystal analysis of the diffraction data revealed that the lowest pressure at which the Y$_3$Fe$_4$H$_{20}$ structure emerges from laser-heating is 60~GPa. The equation of state of Y$_3$Fe$_4$H$_{20}$ was measured by decreasing the pressure down to ambient pressure at room temperature (Fig.~\ref{fig:3}). Remarkably, Y$_3$Fe$_4$H$_{20}$ did not decompose upon decompression. The phase remained orthorhombic with space group $Cmcm$ from 90 to 0~GPa. The single-crystal diffraction data quality decreased progressively with decompression, as expected. In one case, a distortion of single crystal domains towards a lower-symmetry monoclinic phase occurred below 10~GPa, with a more apparent effect on the angle $\beta$ and the lattice parameter c, as shown in Fig.~\ref{fig:3} and Supplementary Material. We fitted the $V(P)$ data points with a Vinet-type equation and obtained an ambient pressure volume $V_0$=703.3~(7.7)~\AA$^3$, a bulk modulus $K_0$=93.0~(12.7)~GPa and its pressure derivative $K^{'}_{0}$=5.8~(0.5) (numbers between parentheses represent 95\% confidence intervals). 

The compression curve and lattice parameters obtained from DFT structural optimizations for Y$_3$Fe$_4$H$_{20}$ are in excellent agreement with our experimental data, as seen in Fig.~\ref{fig:3}. The relative error on the volume is 1.4\% corresponding to less than 0.5\% on the lattice parameters. 
In order to explain the experimentally observed distortion at 0~GPa, we took the conventional (108-atom) unit cell, as obtained in the $Cmcm$ space group at 0~GPa, applied random displacements of a few 0.01~\AA\ to each atomic position along all three directions, set the $\beta$ angle to 93$^{\circ}$, and fully re-optimized the system with DFT calculations. This makes sense since, experimentally, the 108-atom unit cell is preserved in the distorted system, showing that, if the distortion is induced by a phonon dynamical instability, this instability necessarily takes place at the $\Gamma$ point of the Brillouin zone associated with the conventional unit cell.
However, the calculation converged back to the undistorted structure with the $Cmcm$ space group. This suggests that the experimentally observed distortion does not arise from a dynamical instability, but rather from 
non-hydrostatic pressure conditions occurring during decompression of the diamond anvil cell, although other causes cannot be ruled out.

\subsection*{Structural similarities with inosilicate}
The electron density distribution (Fig.~\ref{fig:4}) and electron localization function (ELF, see the Supplementary Material) were calculated to infer the chemical bonding in Y$_3$Fe$_4$H$_{20}$. They both indicate high electronic density on metallic atoms, specially suggesting an electronic localization around Fe. A covalent bond between Fe and H can also be seen in Fig.~\ref{fig:4} for an isovalue of $\sim 0.1$. The isosurface of electronic density distribution at 0.1 is shown within the primitive cell in Fig.~\ref{fig:4}(a).
The Y atoms do not share much electron density with their neighbors while the Fe and H atoms do. Yttrium atoms are surrounded by zero electronic density, while the hydrogen atoms around yttrium are bonded with each other (with a low value 0.08) and with Fe (0.13). This suggests that Fe and H are rather covalently bonded while Y atoms rather form ionic bonds with the Fe-H sublattice. However, the hydrogen cages around the Y atoms differ from those in clathrate superhydrides, where the ELF values are very high. The Fe atoms are surrounded by 8~H atoms, forming [FeH$_8$] molecular complexes. However, in contrast to LaBeH$_{8}$ which can be considered as a complex hydride since it is formed by a rocksalt structure of La$^{2+}$ and BeH$_{8}^{2-}$ units, here the [FeH$_8$] cubic molecular units are bonded together edge-to-edge by hydrogen atoms. These 'chains' of [FeH$_8$] are tied to Y by ionic bonding. This structure can be considered analogous to some found in silicate compounds \cite{Castroviejo2023,Day2020,Verma2023}: two types of hydrogen can be distinguished, non-bridging hydrogen which are bonded to a single Fe and pointing towards Y (Fig.~\ref{fig:4}(c)), indicating an ionic bonding between H and Y; and bridging hydrogen which are bonded with two Fe (Fe-H-Fe), and participate in forming the chains. The chains units contain three polyhedra of 8~H between two cubes of 8~H, which are repeated in the lattice. This allows us to identify a new class of hydrogen-rich compounds: silicate-like superhydrides, with chains of hydrido-complex and counterions.


\section*{Discussion}

Y$_3$Fe$_4$H$_{20}$ exhibits metastability at ambient pressure both in a hydrogen environment and in room air. To assess the compound's metastability, we conducted a series of tests. Samples of run2 were observed through X-ray diffraction during decompression down to ambient pressure and for several hours at ambient conditions. Evidence of metastability was established after more than a day ($\sim$ 30 hours) as shown in Fig.~\ref{fig:5}. Another sample from run1 was retained and monitored through X-ray diffraction after a period of three months. A volume reduction of 7\% was observed, which we attribute to hydrogen desorption, showing a limit of the metastability. Notably, both orthorhombic and monoclinic crystallites are observed after 30 hours at ambient pressure. The remarkable metastability of this superhydride at lower and ambient pressure is also confirmed by first-principles calculations. As stated earlier, the formation of Y$_3$Fe$_4$H$_{20}$ is energetically favorable with respect to simpler compounds in all cases, except at 0~GPa in the case of certain specific reactions, which appears to be consistent with expectations (see the Supplementary Material). 

The nearest H-H distance in this material is approximately 1.33~\AA, which is similar to that observed in FeH$_5$~\cite{Pepin2017} at 90~GPa. We calculated the H-H distances in the $Cmcm$ structure down to ambient pressure. The pressure evolution of the nearest H-H distance (shown in the Supplementary Material) follows a similar trend as in FeH$_5$. That nearest neighbour superhydride criteria remains valid down to ambient pressure. 
However, unlike FeH$_5$, this material remains metastable at ambient pressure. This metastability can be attributed to the formation of the silicate-like chains of [FeH$_8$] units. The structural stability of the Y$_3$Fe$_4$H$_{20}$ superhydride under ambient conditions for several hours is a clear indication that certain structural features of superhydrides, typically stabilized in the 100~GPa range, can persist in a metastable state at ambient pressure.
The XRD single-crystal structural refinement of the reaction products obtained in a laser-heated DAC is presented here as a promising approach for exploring ternary hydrides. This method was inspired by its success in the search of polynitrogen compounds, leading to the discovery of structures beyond the current capabilities of random \textit{ab initio} searches~\cite{Laniel2023}. The structure we predicted has a 54-atom primitive unit cell, meaning that finding Y$_3$Fe$_4$H$_{20}$ through calculations alone would have required a computationally expensive large-scale exploration with large unit cells. Simultaneously, high-pressure experimental XRD alone cannot provide the arrangement of the H-sublattice. Therefore, a dialogue between experiment and computation is essential for the discovery and study of novel superhydrides. 

It has recently been noted that most prototype clathrate-like structures adopted by binary superhydrides can be found in zeolite databases~\cite{Jiang2025}, as they share similar 3D pore frameworks. Following this logic, the results obtained here indicate that aluminosilicate structures could also serve as another promising database for identifying potential ternary superhydrides. In a first step, the present structure can now serve as a template for optimizing compound properties through chemical substitution. As a starting example, substituting Fe by Ni results in a possible Y$_3$Ni$_4$H$_{20}$ superhydride. DFT calculations on this compound show a slightly enhanced contribution of hydrogen-projected density of states at the Fermi level (see the Supplementary Material), making it a more promising candidate for superconductivity. 
%

\section*{Methods}
\subsection*{Experimental details}

The sample configuration and the synchrotron X-ray diffraction measurements were similar to those used in previous studies of compounds formed in the Fe-H~\cite{Pepin2017} or U-H~\cite{Guigue2020} systems at high pressures. Experiments were performed on 3-8~$\mu m$ YFe$_2$H$_{4.2}$ cristalline samples loaded in diamond anvil cells (DAC) with hydrogen as pressure transmitting medium (PTM). The YFe$_2$H$_{4.2}$ hydrogenated compound was chosen to perform hydrogenation under high-hydrogen pressure to prevent any hydrogen-induced amorphization or decomposition into YH$_3$ and $\epsilon$-Fe.
Rhenium gaskets were used, protected with a 1.5-6~$\mu m$ gold layer to prevent hydrogen diffusion. Pressure was measured using the equation of state of gold in the coating close to the sample in the experimental chamber of the DAC. The uncertainty in pressure using this gold pressure gauge is about 2\%. 

There were three runs of compression/decompression of the sample in hydrogen PTM, designated as run1, run2 and run3. 
In run1, two samples \#S1 and \#S2 were loaded, both about 8$\mu m$ in diameter, in a DAC with 100$\mu m$ diamond culets. The DAC was compressed up to 83~GPa and offline laser heating was performed on both samples. After the synthesis, we followed its evolution with pressure by first increasing the pressure up to 90~GPa and then decreasing it down to 0~GPa, at ambient temperature. Both samples were followed in order to check the repeatability. One sample recovered at ambient pressure was kept for three months to evaluate its long-term metastability. 

In run2, two samples \#S3 and \#S4 about $6\mu m$ and $10\mu m$ in diameter were loaded in a DAC with 300~$\mu m$ diamond culets. The samples were compressed up to 60~GPa accompanied by offline laser heating at three different pressures: 30, 45 and 60~GPa. The aim of this run was to determine the lowest synthesis pressure of the new compound, and then to evaluate the hydrogen desorption speed of Y$_3$Fe$_4$H$_{20}$ at ambient conditions after pressure decompression. The lattice volume of the resulting samples after pressure decompression was monitored during 30~hours without noticeable change.

In run3, a single bigger sample \#S5 about 14$\mu m$ in diameter was loaded in a DAC with 100~$\mu m$ diamond culets. Y$_3$Fe$_4$H$_{20}$ was synthesized by laser heating at 80~GPa. XRD measurements, including mapping, were performed to check the composition and homogeneity of the synthesized sample. Further studies on the metastability, and retrieval of the sample, were prevented by the sudden closing of the gasket at 40~GPa during decompression. 

Offline laser heating was performed with an ytterbium fiber laser with a 1070~nm wavelength. We maintained the heating temperature below 1500~K to prevent hydrogen diffusion in the diamond anvils. The XRD data were collected at the ESRF ID15B and ID27 beamlines on accepted proposals HC-5070 and HC-5453. XRD measurements were performed with a monochromatic wavelength of either 0.3738\AA~or 0.411\AA~and diffraction data were acquired on an EIGER2 X CdTe 9M flat-panel detector. The X-ray beam spot size was about 1.5$\times$1.5$\mu m^2$. A single crystal of vanadinite Pb$_5$(VO$_4$)$_3$Cl (space group $P63/m$, a =10.299 Å, c = 7.308 Å) was used as calibration standard for refinement of the instrument model of the diffractometer. Preliminary analysis was performed using Dioptas software~\cite{Prescher2015}. Single crystal analysis was performed by combining the use of the CrysAlisPro~\cite{Crysalis}, DAFi~\cite{Aslandukov2022} and Jana2006~\cite{Petricek2014} softwares.

\subsection*{Computational details}

DFT calculations were performed with the ABINIT code~\cite{Gonze2016}, within the Projector-Augmented Wave (PAW) formalism~\cite{Torrent2008}. The exchange-correlation energy functional used throughout this work was the generalized gradient approximation as formulated by Perdew, Burke, and Ernzerhof (GGA-PBE)~\cite{Perdew1996}. We took the PAW atomic datasets for Y and Fe from the Jollet-Torrent-Holzwarth (JTH) table~\cite{Jollet2014}. For H, we used two PAW atomic datasets: the one from the JTH table (PAW radius $\sim$ 1.0~bohr) and another one with a smaller PAW radius of 0.8~bohr.

We explored different hydrogen content by randomly searching hydrogen distributions within the interstitial sites of the orthorhombic $Cmcm$ structure of Y$_3$Fe$_4$H$_x$. Indeed, single crystal analysis enables to resolve the metallic structure of the compound, but not its hydrogen content. Experimental data analysis gave essential information to guide and constrain the random search: the space group $Cmcm$, the cell volume 500\AA$^3$/4f.u., the lattice parameters and the atomic positions of Y and Fe atoms at 83~GPa. Y$_3$Fe$_4$H$_x$ can be seen as a mix of yttrium and iron superhydrides: YH$_3$, FeH$_2$ and FeH$_3$ at 83~GPa. As a mix of YH$_3$ and FeH$_2$, the hydrogen content of the ternary compound Y$_3$Fe$_4$H$_x$ would be 17~H per formula unit. As a mix of YH$_3$ and FeH$_3$, the hydrogen content of the ternary superhydrides Y$_3$Fe$_4$H$_x$ would be 21~H per formula unit. Thus, we only explored hydrogen stoichiometries between Y$_3$Fe$_4$H$_{17}$ and Y$_3$Fe$_4$H$_{21}$.

We started from the conventional unit cell of Y$_3$Fe$_4$ found experimentally (4 formula units), and then randomly drew the hydrogen positions in this cell between 17H/f.u. to 21H/f.u., within the Wyckoff positions of the $Cmcm$ space group. In all generated structures, iron and yttrium positions were fixed using the experimental positions, i.e. the following Wyckoff sites for Y: 4c and 8f, for Fe: 4b, 8f and 4c.
We used the conventional unit cell for the random generation of structures, rather than the primitive one, because $Cmcm$ Wyckoff sites are traditionally described in the conventional cell.

For the random generation of hydrogen positions, we used the random number generator of R. Chandler and P. Northrup~\cite{rngChandler}. We proceeded as follows: within the conventional unit cell, and for a given hydrogen stoichiometry, a determined number of hydrogen positions had to be randomly drawn among the Wyckoff sites of the $Cmcm$ space group. While this number was not reached, a Wyckoff site was randomly drawn among the possible ones, along with the necessary x, y, z positions. This step was repeated until the exact amount of hydrogen was reached (the multiplicity of each site was taken into account). Among the 8 possible Wyckoff sites of the $Cmcm$ space group, 8d can be drawn at most once because it has determined positions. 4b also has determined positions but is already occupied by a Fe. 
4a is excluded since it implies unphysical arrangement of atoms.
The random configurations having atoms too close to each other were systemetically rejected. The number of Wyckoff sites randomly drawn also depends on the explored hydrogen stoichiometries.

In order to gain computational time, the generated structures were then reduced to primitive cells with 2 f.u., and finally fully structurally optimized (all atomic positions including Fe and Y, and lattice parameters, within the $Cmcm$ space group). The structural optimizations at the random searching stage were performed using the BFGS algorithm as implemented in ABINIT, with the following parameters: the Brillouin zone associated with the simulation cell was sampled by a $4\times4\times2$ {k}-point mesh, a plane-wave cutoff of 15 Hartree was employed, and the optimization criterion on atomic forces was 2.0$\times10^{-3}$ Hartree/bohr, with a number of ionic steps typically limited to 75 steps. 
The $4\times4\times2$ k-point mesh provides a numerical precision better than 1 meV/atom (typically a few 0.1 meV/atom) on the total energy of Y$_3$Fe$_4$H$_{20}$, while the 15 Hartree plane-wave cut-off provides a precision of $\sim$ 1 meV/atom on the formation energy of Y$_3$Fe$_4$H$_{20}$.
This method is inspired from the \textit{ab initio} random searching (AIRSS) algorithm~\cite{Pickard2011}, with the exception that the underlying crystal structure is imposed by the experimentally determined $Cmcm$-Y$_3$Fe$_4$H$_x$, and that only the positions of hydrogen atoms are randomly searched throughout available Wyckoff sites of the space group. 

At the end of each structural optimization, the final enthalpy was computed, which allowed to identify the most stable structure by comparing all the enthalpies obtained for a given hydrogen stoichiometry. The number of structural optimizations performed for each considered hydrogen stoichiometry was typically varying between 1000 and 2000. Spin-polarization was not taken into account in this random search process.

Candidate structures were then selected as those having the lowest enthalpy. The identified candidates (around 5 per considered hydrogen stoichiometry) were then re-optimized using the BFGS algorithm with more stringent numerical parameters: a plane-wave cut-off of 25 Hartree, 8$\times$8$\times$4~{k}-point mesh to sample the Brillouin zone associated with the simulation cell, and an optimization criterion on the atomic forces of 1.0~$\times$~10$^{-5}$ Hartree/bohr (and 1.0~$\times$~10$^{-7}$ Hartree/Bohr$^3$ on the stress tensor components).

The calculation of the formation enthalpies of Y$_3$Fe$_4$H$_{17}$ requires to have the enthalpies of Y$_3$Fe$_4$ and H$_2$ under pressure. We have thus structurally optimized these two systems under pressure, using the same criteria, and dense hydrogen in its $C2/c$-$32$ structure. However, there is to our knowledge no evidence that the Y$_3$Fe$_4$ alloy exists at this pressure or below. Thus we chose to use Y$_3$Fe$_4$H$_{17}$ as the reference for plotting the convex-hull and not Y$_3$Fe$_4$. Experimental excess hydrogen ensures that the thermodynamically favored structure is the one lying on the convex-hull nearest H$_{2}$, which in this case is Y$_3$Fe$_4$H$_{20}$.

The dynamical stability of the predicted crystal structure Y$_3$Fe$_4$H$_{20}$ superhydride was then tested at 90~GPa by computing the phonon dispersion curves. These calculations were performed within the framework of the density-functional perturbation theory (DFPT), as implemented in ABINIT~\cite{Gonze2016}. For these calculations, the structure was first fully re-optimized. We used a 4$\times$4$\times$2 $\Gamma$-centered q-point mesh in DFPT calculations to sample the Brillouin zone. 

We then calculated the formation enthalpies at several pressures from 90 to 0~GPa and compared it to that of the decomposition compounds to ensure Y$_3$Fe$_4$H$_{20}$ stability. The selected decomposition compounds and the results are listed in the Supplementary Material.

Finally, we computed the electronic density distribution and the electron localization functions (ELF) within the norm-conserving formalism (ELF is not available in PAW formalism in ABINIT) to evaluate the nature of hydrogen-hydrogen and metal-hydrogen bonds.

The study of the Laves phase hydrides YFe$_2$H$_6$ and YFe$_2$H$_7$~\cite{Causse2025} highlighted the importance of taking into account spin-polarization and Hubbard corrections on iron atoms. The compression curves of Y$_3$Fe$_4$H$_{20}$ calculated under various approximations are compared to experiment and GGA-PBE calculations (see the Supplementary Material). Interestingly, we see that within the GGA-PBE, taking into account spin degrees of freedom initialized with an hypothetical ferromagnetic order in Y$_3$Fe$_4$H$_{20}$ always reverts to the non-magnetic state. Thus it gives exactly the same results as the calculated compression curve without spin degrees of freedom. 
Including electronic correlations, by adding a Hubbard correction on the Fe $d$-orbitals within the GGA-PBE+$U$ framework ($U$=2.6 eV and $J$=0.9 eV~\cite{Dewaele2023}) provides close results, with however a slightly ferromagnetic system, the magnetic moment per Fe atom decreasing from $\sim$0.75~$\mu_B$ at 0~GPa to $\sim$0.25~$\mu_B$ at $\sim$70~GPa, and then to zero at higher pressure.


\section*{Supplementary Material}

See the Supplementary Material for additional information.

\section*{Acknowledgements}

We acknowledge the European Synchrotron Radiation Facility (ESRF) for provision of synchrotron radiation facilities  under proposal numbers HC-5070~\cite{ProposalHC5070} and HC-5453~\cite{ProposalHC5453} and we would like to thank G.~Garbarino for assistance and support in using beamlines ID15B and ID27. We thank V.~Schmidt for his help on the DAC preparations. We thank F.~Occelli for his help with H$_2$ loading. We thank V.~Paul-Boncour for synthesizing the precursor sample, fruitful discussions and comments. We thank D.~Laniel for his valuable comments on single-crystal data analysis. The crystal drawings have been made with the VESTA software~\cite{vesta2011}.

\section*{Author declarations}
\subsection*{Conflict of Interest}
The authors have no conflicts to disclose.

\subsection*{Author Contributions}
G.G. and P.L. designed research; M.C. and L.T. performed XRD measurements and structural refinement; M.C. and G.G. performed \textit{ab initio} calculations; All authors analyzed data and wrote the manuscript.

\section*{Data Availability}
The data that support the findings of this study are available from the corresponding author upon reasonable request.

\clearpage

\section*{References}

\bibliographystyle{unsrt}
\bibliography{biblio.bib}

\begin{thebibliography}{10}

\bibitem{Zurek2009}
Eva Zurek, Roald Hoffmann, N.~W. Ashcroft, Artem~R. Oganov, and Andriy~O.
  Lyakhov.
\newblock A little bit of lithium does a lot for hydrogen.
\newblock {\em Proceedings of the National Academy of Sciences},
  106(42):17640--17643, 2009.

\bibitem{FloresLivas2020}
José~A. Flores-Livas, Lilia Boeri, Antonio Sanna, Gianni Profeta, Ryotaro
  Arita, and Mikhail Eremets.
\newblock A perspective on conventional high-temperature superconductors at
  high pressure: Methods and materials.
\newblock {\em Physics Reports}, 856:1--78, 2020.
\newblock A perspective on conventional high-temperature superconductors at
  high pressure: Methods and materials.

\bibitem{Du2022}
Mingyang Du, Wendi Zhao, Tian Cui, and Defang Duan.
\newblock Compressed superhydrides: the road to room temperature
  superconductivity.
\newblock {\em Journal of Physics: Condensed Matter}, 34(17):173001, 2022.

\bibitem{Sun2023}
Ying Sun, Xin Zhong, Hanyu Liu, and Yanming Ma.
\newblock {Clathrate metal superhydrides under high-pressure conditions:
  enroute to room-temperature superconductivity}.
\newblock {\em National Science Review}, 11(7):nwad270, 10 2023.

\bibitem{Pepin2017}
C.~M. P{\'e}pin, G.~Geneste, A.~Dewaele, M.~Mezouar, and P.~Loubeyre.
\newblock Synthesis of feh5: A layered structure with atomic hydrogen slabs.
\newblock {\em Science}, 357(6349):382--385, 2017.

\bibitem{Geballe2018}
Zachary~M. Geballe, Hanyu Liu, Ajay~K. Mishra, Muhtar Ahart, Maddury
  Somayazulu, Yue Meng, Maria Baldini, and Russell~J. Hemley.
\newblock Synthesis and stability of lanthanum superhydrides.
\newblock {\em Angewandte Chemie International Edition}, 57(3):688--692, 2018.

\bibitem{Somayazulu2019}
Maddury Somayazulu, Muhtar Ahart, Ajay~K. Mishra, Zachary~M. Geballe, Maria
  Baldini, Yue Meng, Viktor~V. Struzhkin, and Russell~J. Hemley.
\newblock Evidence for superconductivity above 260 k in lanthanum superhydride
  at megabar pressures.
\newblock {\em Phys. Rev. Lett.}, 122:027001, 2019.

\bibitem{Drozdov2019}
A.~P. Drozdov, P.~P. Kong, V.~S. Minkov, S.~P. Besedin, M.~A. Kuzovnikov,
  S.~Mozaffari, D.~A. Balicas, M.~Tkacz, and M.~I. Eremets.
\newblock Superconductivity at 250 k in lanthanum hydride under high pressures.
\newblock {\em Nature}, 569:528, 2019.

\bibitem{Causse2023}
Ma\'elie Causs\'e, Gr\'egory Geneste, and Paul Loubeyre.
\newblock Superionicity of ${\mathrm{h}}^{\ensuremath{\delta}\ensuremath{-}}$
  in ${\mathrm{lah}}_{10}$ superhydride.
\newblock {\em Phys. Rev. B}, 107:L060301, 2023.

\bibitem{Sun2019}
Ying Sun, Jian Lv, Yu~Xie, Hanyu Liu, and Yanming Ma.
\newblock Route to a superconducting phase above room temperature in
  electron-doped hydride compounds under high pressure.
\newblock {\em Phys. Rev. Lett.}, 123:097001, August 2019.

\bibitem{Semenok2021}
Dmitrii~V. Semenok, Ivan~A. Troyan, Anna~G. Ivanova, Alexander~G. Kvashnin,
  Ivan~A. Kruglov, Michael Hanfland, Andrey~V. Sadakov, Oleg~A. Sobolevskiy,
  Kirill~S. Pervakov, Igor~S. Lyubutin, Konstantin~V. Glazyrin, Nico Giordano,
  Denis~N. Karimov, Alexander~L. Vasiliev, Ryosuke Akashi, Vladimir~M. Pudalov,
  and Artem~R. Oganov.
\newblock Superconductivity at 253 k in lanthanum–yttrium ternary hydrides.
\newblock {\em Materials Today}, 48:18--28, 2021.

\bibitem{Bi2022}
Jingkai Bi, Yuki Nakamoto, Peiyu Zhang, Katsuya Shimizu, Bo~Zou, Hanyu Liu,
  Mi~Zhou, Guangtao Liu, Hongbo Wang, and Yanming Ma.
\newblock Giant enhancement of superconducting critical temperature in
  substitutional alloy (la,ce)h9.
\newblock {\em Nature Communications}, 13(1):5952, 2022.

\bibitem{Zhang2022}
Zihan Zhang, Tian Cui, Michael~J. Hutcheon, Alice~M. Shipley, Hao Song,
  Mingyang Du, Vladimir~Z. Kresin, Defang Duan, Chris~J. Pickard, and Yansun
  Yao.
\newblock Design principles for high-temperature superconductors with a
  hydrogen-based alloy backbone at moderate pressure.
\newblock {\em Phys. Rev. Lett.}, 128:047001, January 2022.

\bibitem{Song2023}
Yinggang Song, Jingkai Bi, Yuki Nakamoto, Katsuya Shimizu, Hanyu Liu, Bo~Zou,
  Guangtao Liu, Hongbo Wang, and Yanming Ma.
\newblock Stoichiometric ternary superhydride ${\mathrm{labeh}}_{8}$ as a new
  template for high-temperature superconductivity at 110 k under 80 gpa.
\newblock {\em Phys. Rev. Lett.}, 130:266001, June 2023.

\bibitem{Wang2024}
Xiaoxue Wang, Yuqing Ding, and Hui Wang.
\newblock First-principles study of the dynamics in face-centered cubic ceh9
  and ceh10 under high pressure.
\newblock {\em Chinese Journal of High Pressure Physics}, 38(2):1--7, April
  2024.

\bibitem{Dolui2024}
Kapildeb Dolui, Lewis~J. Conway, Christoph Heil, Timothy~A. Strobel, Rohit~P.
  Prasankumar, and Chris~J. Pickard.
\newblock Feasible route to high-temperature ambient-pressure hydride
  superconductivity.
\newblock {\em Phys. Rev. Lett.}, 132:166001, April 2024.

\bibitem{Sanna2024}
Antonio Sanna, Tiago F.~T. Cerqueira, Yue-Wen Fang, Ion Errea, Alfred Ludwig,
  and Miguel A.~L. Marques.
\newblock Prediction of ambient pressure conventional superconductivity above
  80 k in hydride compounds.
\newblock {\em npj Computational Materials}, 10(1):44, February 2024.

\bibitem{Hansen2024}
Mads~F. Hansen, Lewis~J. Conway, Kapildeb Dolui, Christoph Heil, Chris~J.
  Pickard, Anna Pakhomova, Mohammed Mezouar, Martin Kunz, Rohit~P. Prasankumar,
  and Timothy~A. Strobel.
\newblock Synthesis of mg2irh5: A potential pathway to high-$t_c$ hydride
  superconductivity at ambient pressure, 2024.

\bibitem{Aslandukov2022}
Andrey Aslandukov, Matvii Aslandukov, Natalia Dubrovinskaia, and Leonid
  Dubrovinsky.
\newblock {{\it Domain Auto Finder} ({\it DAFi}) program: the analysis of
  single-crystal X-ray diffraction data from polycrystalline samples}.
\newblock {\em Journal of Applied Crystallography}, 55(5):1383--1391, Oct 2022.

\bibitem{Pepin2015}
Charles Pépin, Paul Loubeyre, Florent Occelli, and Paul Dumas.
\newblock Synthesis of lithium polyhydrides above 130 gpa at 300 k.
\newblock {\em Proceedings of the National Academy of Sciences},
  112(25):7673--7676, 2015.

\bibitem{Kong2021}
Panpan Kong, Vasily~S. Minkov, Mikhail~A. Kuzovnikov, Alexander~P. Drozdov,
  Stanislav~P. Besedin, Shirin Mozaffari, Luis Balicas, Fedor~Fedorovich
  Balakirev, Vitali~B. Prakapenka, Stella Chariton, Dmitry~A. Knyazev, Eran
  Greenberg, and Mikhail~I. Eremets.
\newblock Superconductivity up to 243 k in the yttrium-hydrogen system under
  high pressure.
\newblock {\em Nature Communications}, 12(1):5075, August 2021.

\bibitem{Causse2025}
Maélie Caussé, Grégory Geneste, Loïc Toraille, Bastien Guigue,
  Jean-Baptiste Charraud, Valérie Paul-Boncour, and Paul Loubeyre.
\newblock Synthesis of laves phase hydrides yfe2h6 and yfe2h7 at high pressure:
  Reaching a limit of interstitial hydrogen uptake.
\newblock {\em Journal of Alloys and Compounds}, 1010:177392, 2025.

\bibitem{Spektor2020}
Kristina Spektor, Wilson~A. Crichton, Stanislav Filippov, Sergei~I. Simak,
  Andreas Fischer, and Ulrich Häussermann.
\newblock Na3feh7 and na3coh6: Hydrogen-rich first-row transition metal
  hydrides from high pressure synthesis.
\newblock {\em Inorganic Chemistry}, 59(22):16467--16473, 2020.
\newblock PMID: 33141575.

\bibitem{Crysalis}
{Rigaku Oxford Diffraction}.
\newblock {{{CrysAlisPro Software System}}, {{Version}} 1.171 ({{Rigaku
  Corporation}}, {{Oxford}}, {{UK}}, 2021)}, {2021}.

\bibitem{Petricek2014}
Vaclav Petricek, Michal Dusek, and Lukas Palatinus.
\newblock Crystallographic computing system jana2006: General features.
\newblock {\em Zeitschrift fur Kristallographie - Crystalline Materials},
  229(5):345--352, 2014.

\bibitem{Pickard2011}
Chris~J Pickard and R~J Needs.
\newblock Ab initio random structure searching.
\newblock {\em Journal of Physics: Condensed Matter}, 23(5):053201, January
  2011.

\bibitem{Castroviejo2023}
Ricardo Castroviejo.
\newblock {\em Silicates}, pages 819--844.
\newblock Springer International Publishing, Cham, 2023.

\bibitem{Day2020}
Maxwell~C. Day and Frank~C. Hawthorne.
\newblock A structure hierarchy for silicate minerals: chain, ribbon, and tube
  silicates.
\newblock {\em Mineralogical Magazine}, 84(2):165–244, 2020.

\bibitem{Verma2023}
Pramod~K. Verma.
\newblock {\em Inosilicates}, pages 241--276.
\newblock Springer Nature Switzerland, Cham, 2023.

\bibitem{Laniel2023}
Dominique Laniel, Florian Trybel, Yuqing Yin, Timofey Fedotenko, Saiana
  Khandarkhaeva, Andrey Aslandukov, Georgios Aprilis, Alexei~I. Abrikosov,
  Talha Bin~Masood, Carlotta Giacobbe, Eleanor~Lawrence Bright, Konstantin
  Glazyrin, Michael Hanfland, Jonathan Wright, Ingrid Hotz, Igor~A. Abrikosov,
  Leonid Dubrovinsky, and Natalia Dubrovinskaia.
\newblock Aromatic hexazine [n(6)](4-) anion featured in the complex structure
  of the high-pressure potassium nitrogen compound k(9)n(56).
\newblock {\em Nature chemistry}, 15:641--646, May 2023.

\bibitem{Jiang2025}
Bowen Jiang, Xiaoshan Luo, Ying Sun, Xin Zhong, Jian Lv, Yu~Xie, Yanming Ma,
  and Hanyu Liu.
\newblock Data-driven search for high-temperature superconductors in ternary
  hydrides under pressure.
\newblock {\em Phys. Rev. B}, 111:054505, Feb 2025.

\bibitem{Guigue2020}
Bastien Guigue, Grégory Geneste, Brigitte Leridon, and Paul Loubeyre.
\newblock {An x-ray study of palladium hydrides up to 100 GPa: Synthesis and
  isotopic effects}.
\newblock {\em Journal of Applied Physics}, 127(7):075901, 02 2020.

\bibitem{Prescher2015}
Clemens Prescher and Vitali~B. Prakapenka.
\newblock Dioptas: a program for reduction of two-dimensional x-ray diffraction
  data and data exploration.
\newblock {\em High Pressure Research}, 35(3):223--230, 2015.

\bibitem{Gonze2016}
X.~Gonze, F.~Jollet, F.~{Abreu Araujo}, D.~Adams, B.~Amadon, T.~Applencourt,
  C.~Audouze, J.-M. Beuken, J.~Bieder, A.~Bokhanchuk, E.~Bousquet, F.~Bruneval,
  D.~Caliste, M.~Côté, F.~Dahm, F.~{Da Pieve}, M.~Delaveau, M.~{Di Gennaro},
  B.~Dorado, C.~Espejo, G.~Geneste, L.~Genovese, A.~Gerossier, M.~Giantomassi,
  Y.~Gillet, D.R. Hamann, L.~He, G.~Jomard, J.~{Laflamme Janssen}, S.~{Le
  Roux}, A.~Levitt, A.~Lherbier, F.~Liu, I.~Lukacević, A.~Martin, C.~Martins,
  M.J.T. Oliveira, S.~Poncé, Y.~Pouillon, T.~Rangel, G.-M. Rignanese, A.H.
  Romero, B.~Rousseau, O.~Rubel, A.A. Shukri, M.~Stankovski, M.~Torrent, M.J.
  {Van Setten}, B.~{Van Troeye}, M.J. Verstraete, D.~Waroquiers, J.~Wiktor,
  B.~Xu, A.~Zhou, and J.W. Zwanziger.
\newblock Recent developments in the abinit software package.
\newblock {\em Computer Physics Communications}, 205:106--131, 2016.

\bibitem{Torrent2008}
Marc Torrent, François Jollet, François Bottin, Gilles Zérah, and Xavier
  Gonze.
\newblock Implementation of the projector augmented-wave method in the abinit
  code: Application to the study of iron under pressure.
\newblock {\em Computational Materials Science}, 42(2):337--351, 2008.

\bibitem{Perdew1996}
John~P. Perdew, Kieron Burke, and Matthias Ernzerhof.
\newblock Generalized gradient approximation made simple.
\newblock {\em Phys. Rev. Lett.}, 77:3865--3868, 1996.

\bibitem{Jollet2014}
François Jollet, Marc Torrent, and Natalie Holzwarth.
\newblock Generation of projector augmented-wave atomic data: A 71 element
  validated table in the xml format.
\newblock {\em Computer Physics Communications}, 185(4):1246--1254, 2014.

\bibitem{rngChandler}
R.~Chandler and P.~Northrup.
\newblock www.ucl.ac.uk/~ucakarc/work/software/randgen.txt.

\bibitem{Dewaele2023}
Agn\`es Dewaele, Bernard Amadon, Alexei Bosak, Volodymyr Svitlyk, and Florent
  Occelli.
\newblock Synthesis of single crystals of $\ensuremath{\epsilon}$-iron and
  direct measurements of its elastic constants.
\newblock {\em Phys. Rev. Lett.}, 131:034101, Jul 2023.

\bibitem{ProposalHC5070}
M.~Causs\'e, L.~Toraille, and F.~Occelli.
\newblock Synthesis of novel yfe2h(x larger than 10) compounds up to 100 gpa: a
  possibility to recover a high content h-storage material at ambient pressure
  [dataset]., 2022.

\bibitem{ProposalHC5453}
M.~Causs\'e, C.M. P\'epin, V.~Schmidt, and L.~Toraille.
\newblock {Direct synthesis of a ternary superhydride, YxFeyHz compound:
  structure and metastability [dataset].}, 2023.

\bibitem{vesta2011}
Koichi Momma and Fujio Izumi.
\newblock {{\it VESTA3} for three-dimensional visualization of crystal,
  volumetric and morphology data}.
\newblock {\em Journal of Applied Crystallography}, 44(6):1272--1276, Dec 2011.

\end{thebibliography}



\clearpage

\begin{Figure}
\centering
\includegraphics[width=1\columnwidth]{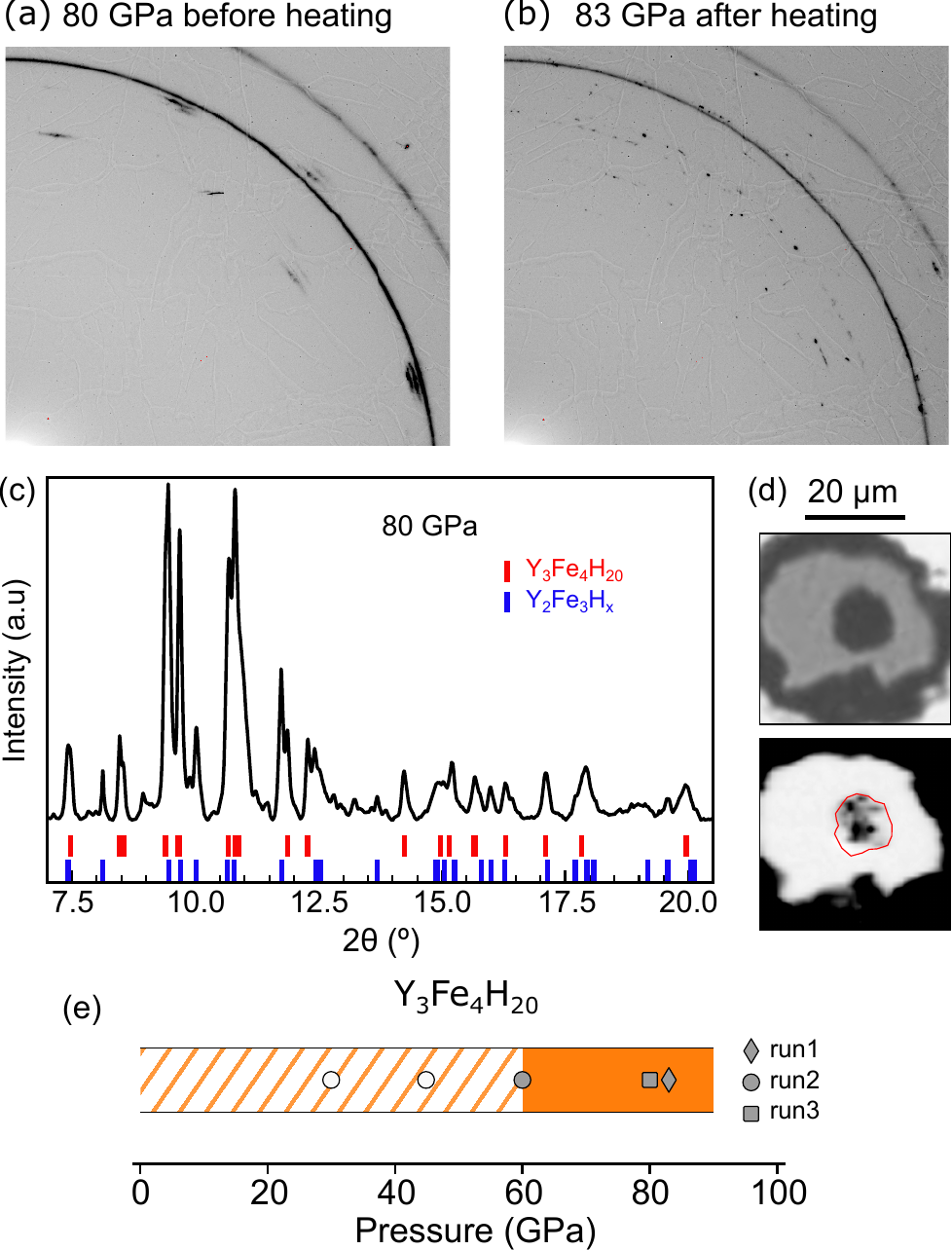}
\caption{High-pressure Y-Fe-H transformation by laser heating. (a) Zoom on a XRD image plate of the precursor Y-Fe-H sample at 83~GPa before laser heating in run1. Continuous rings correspond to the diffraction of the gold liner of the sample chamber. The diffraction spots of the sample show a clear distortion of the original single crystal sample. (b) After laser heating, numerous diffraction spots are visible. Their respective identical $2\theta$ values indicate the presence of multiple similar single crystals oriented in various directions. (c) X-ray diffraction pattern registered at 80~GPa (run3) after heating. The structure resolution through single crystal analysis leads to the indexation of all significant reflections peaks. Two different phases are observed, a dominant phase Y$_3$Fe$_4$H$_x$ (red) and a minor phase Y$_2$Fe$_3$H$_x$ (blue). (d) Top: optical image of the reacted sample in run3, about 15~$\mu m$ in diameter. Bottom: corresponding XRD map analysis, selecting different $2\theta$ values associated to Y$_3$Fe$_4$H$_{20}$. The contour of the sample is outlined in red. The diffraction signature of Y$_3$Fe$_4$H$_{20}$ is distributed almost everywhere in the sample, showing that the sample is quite homogeneous. (e) Pressure domain of the synthesis of Y$_3$Fe$_4$H$_{20}$ under laser heating, different markers identify the three runs. When markers are filled (grey) the phase Y$_3$Fe$_4$H$_{20}$ was observed after laser heating, whereas it was not observed when the markers are empty (white). The orange area indicates the stability domain and the orange dashed-line  shows the metastability region of the phase Y$_3$Fe$_4$H$_{20}$ during decompression to ambient pressure.}\label{fig:1}
\end{Figure}

\clearpage

\begin{Figure}
\centering
\includegraphics[width=1\columnwidth]{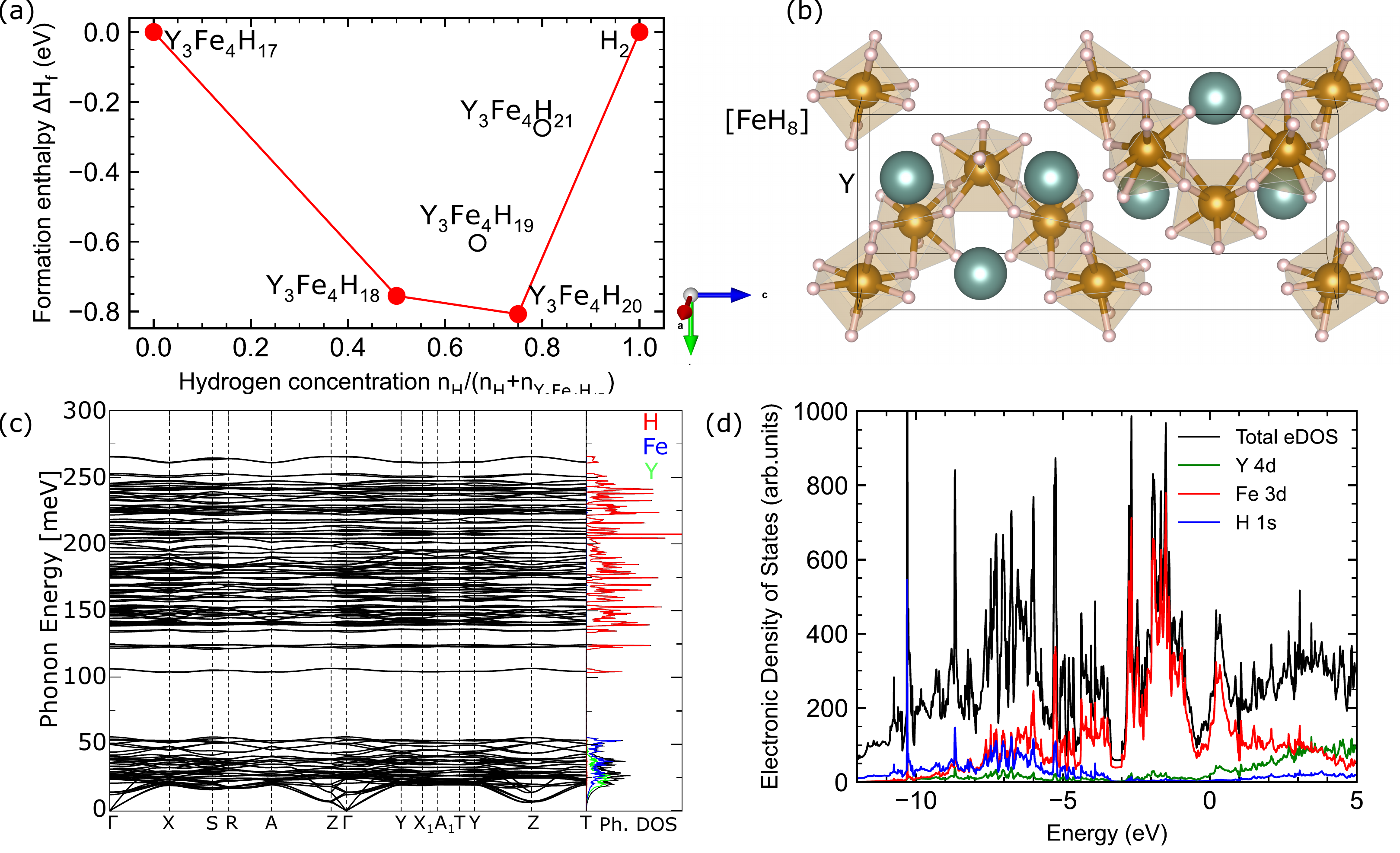}
\caption{Structure, thermodynamic and dynamical stability of Y$_3$Fe$_4$H$_{20}$ (a) Calculated convex hull at 90~GPa for the various $Cmcm$-Y$_3$Fe$_4$H$_x$ compounds. The phases which lie on the convex hull (in red) are thermodynamically stable. The phase Y$_3$Fe$_4$H$_{20}$ is likely to be the experimentally synthesized phase since it was synthesized in excess of hydrogen. (b) Crystal structure of Y$_3$Fe$_4$H$_{20}$ (primitive cell), with the Fe atoms in orange and the Y atoms in blue. The Fe-H covalent bonds are shown. Each Fe atom is surrounded by 8~H forming [FeH$_8$] units. These units are connected edge-to-edge by two hydrogen atoms. (c) GGA-PBE total and projected electronic densities of states of Y$_3$Fe$_4$H$_{20}$ at 90~GPa. The projected 1s-H electrons do not contribute to the Fermi level. The projected Fe-3d electrons are the main contributors at the Fermi level. (d) Results of the DFPT phonons calculations for Y$_3$Fe$_4$H$_{20}$ at 90~GPa: phonon dispersion curves along high-symmetry directions in the Brillouin zone; the green, blue and red density of states correspond to yttrium, iron and hydrogen atoms respectively. }\label{fig:2}
\end{Figure}

\clearpage

\begin{Figure}
\centering
\includegraphics[width=1\columnwidth]{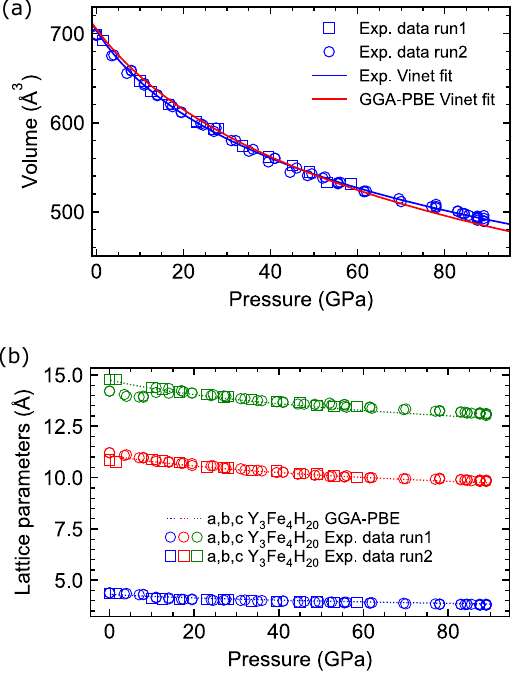}
\caption{Volume and lattice compression curve of  Y$_3$Fe$_4$H$_{20}$ between 0 and 90~GPa. (a) Experimental (in blue) and calculated (in red) equation of state of Y$_3$Fe$_4$H$_{20}$. The curves are fitted with a Vinet equation of state. Data from both samples \#S1 and \#S2 from run1 (square), obtained during decompression, are plotted along with samples \#S3 and \#S4 from run2 (circle). (b) Evolution of the lattice parameters with pressure and comparison between experimental and calculated data. The square and circle symbols follow the same convention as (a). The dotted lines correspond to the lattice parameters calculated with the GGA-PBE functional. The error bars are contained within the symbol sizes.} 
\label{fig:3}
\end{Figure}

\clearpage

\begin{Figure}
\centering
\includegraphics[width=1\columnwidth]{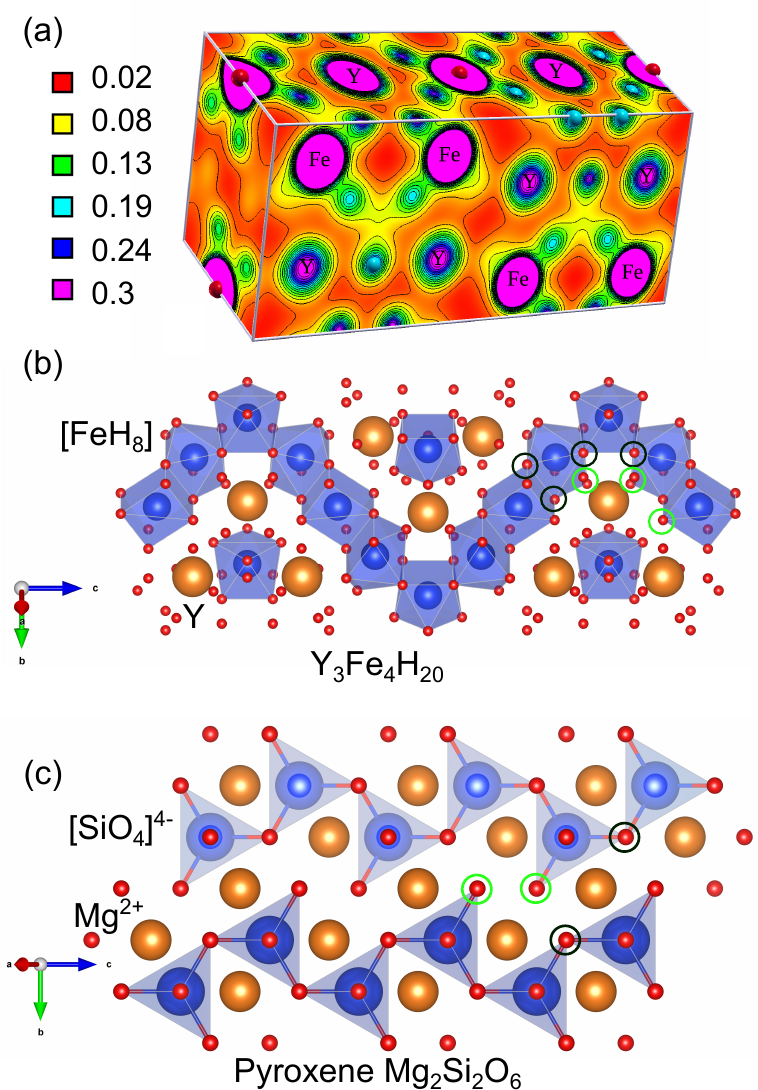}
\caption{Electronic density distribution of Y$_3$Fe$_4$H$_{20}$ and the inosilicate analogy. (a) Electronic density distribution for the orthorhombic primitive cell calculated at 90~GPa. Y atoms are isolated while Fe and H atoms have covalent bonding. Isovalue scale is plotted as a color scale. (b) Crystal structure of Y$_3$Fe$_4$H$_{20}$, with the Fe atoms in blue, the Y atoms in orange and the hydrogen atoms in red. [FeH$_8$] units (blue polyedron) are connected edge-to-edge by two binding hydrogen atoms (black circles), similarly to binding oxygen atoms in silicate. Non-binding hydrogen atoms (green circles) are pointing towards Y cations and do not participate in forming the chains. (c) Representation of an inosilicate (pyroxene) structure with [SiO$_4$]$^{4-}$ tetrahedron (blue) linked by a bridging oxygen (black circles), forming chains, and independent Mg$^{2+}$ cations (in orange). Non-bridging oxygen (green circles) are covalently bonded with one Si atom only and do not participate in forming the chains.} 
\label{fig:4}
\end{Figure}

\clearpage

\begin{Figure}
\centering
\includegraphics[width=1\columnwidth]{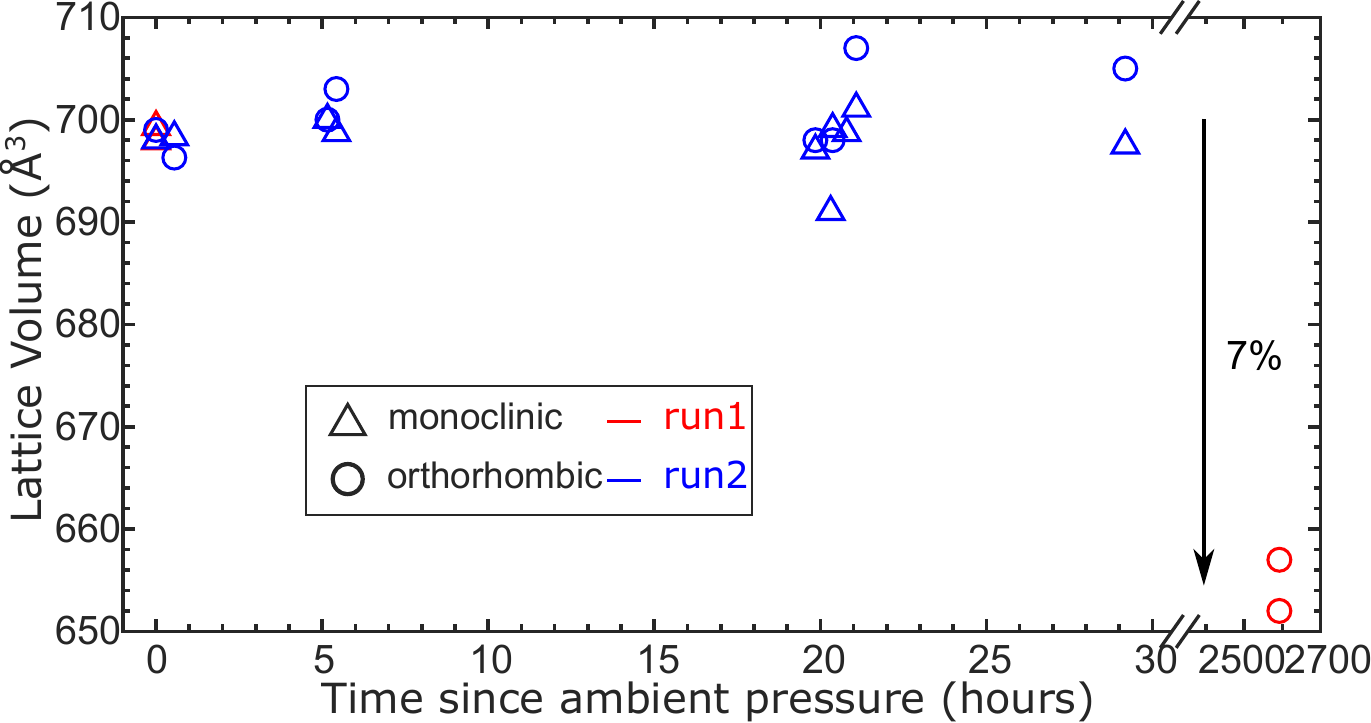}
 \caption{Metastability of the phase Y$_3$Fe$_4$H$_{20}$: evolution of the experimental lattice volume at ambient pressure versus time. Both samples of run1 (in red) at ambient pressure at time zero contained only monoclinic crystallites; three months after (2700 hours) only orthorhombic crystallites were observed with a 7\% reduction of lattice volume. In run2, both monoclinic and orthorhombic crystallites were observed in the sample. No significant structure or lattice volume variation were observed during a 30 hours time period.}
 \label{fig:5}
 \end{Figure}

\end{document}